\documentclass[prl,twocolumn,letterpaper,groupedaddress,amsmath,amssymb,showpacs,aps,english,superscriptaddress,longbibliography]{revtex4-1}

\usepackage{graphicx}
\usepackage{bm,mathrsfs}
\usepackage{amsmath}
\usepackage{amssymb}
\usepackage[colorlinks=true,allcolors=blue]{hyperref}
\usepackage{mathtools}

\makeatletter

\begin{document}

\title{Resonant optical spin initialization and readout of single silicon vacancies in 4H-SiC}

\author{Hunter B. Banks}
\affiliation{NAS-NRC research associate residing at Naval Research Laboratory, Washington, D.C. 20375, USA}
\author{\"{O}ney O. Soykal}
\email[Email: ]{oneysoykal@gmail.com}
\affiliation{Sotera Defense Solutions Inc. residing at Naval Research Laboratory, Washington, D.C. 20375, USA}
\author{Rachael Myers-Ward}
\affiliation{Naval Research Laboratory, Washington, D.C. 20375, USA}
\author{D. Kurt Gaskill}
\affiliation{Naval Research Laboratory, Washington, D.C. 20375, USA}
\author{T. L. Reinecke}
\affiliation{Naval Research Laboratory, Washington, D.C. 20375, USA}
\author{Samuel G. Carter}
\email[Email: ]{sam.carter@nrl.navy.mil}
\affiliation{Naval Research Laboratory, Washington, D.C. 20375, USA}
\begin{abstract}
The silicon monovacancy in 4H-SiC is a promising candidate for solid-state quantum information processing. We perform high-resolution optical spectroscopy on single V2 defects at cryogenic temperatures. We find favorable low temperature optical properties that are essential for optical readout and coherent control of its spin and for the development of a spin-photon interface.  The common features among individual defects include two narrow, nearly lifetime-limited optical transitions that correspond to $m_s{=}\pm 3/2$ and $m_s{=}\pm 1/2$ spin states with no discernable zero-field splitting fluctuations. Initialization and readout of the spin states is characterized by time-resolved optical spectroscopy under resonant excitation of these transitions, showing significant differences between the $\pm 3/2$ and $\pm 1/2$ spin states. These results are well-described by a theoretical model that strengthens our understanding of the quantum properties of this defect.
\end{abstract}

\maketitle

\section{I. Introduction}

Defect centers in wide bandgap materials are candidate systems for use in quantum technologies from sensing to computing, as they harness important advantages of both solid-state and atomic systems \cite{Weber2010, Doherty2013, Aharonovich2016}. Defects can have sharp optical transitions, strong spin-photon coupling, and long coherence times while at the same time allowing for straight-forward nanoscale control and promising scaling.  Nitrogen-vacancy (NV) centers in diamond, owing to their long history and well-understood physics, have become the standard platform for defect-based quantum demonstrations \cite{Childress2006, Doherty2013, Hensen2015}.   Despite their experimental successes, however, NV centers have limitations that have proven difficult to control, from the complexity of nanofabrication to substantial spectral diffusion \cite{Faraon2012, Riedel2017, Siyushev2013}.  These challenges have inspired the search for a new generation of useful defects.

The deep defects of silicon carbide have generated significant interest and shown promise for quantum applications \cite{Castelletto2015, Astakhov2016, Awschalom2018}.  The most immediate and apparent advantages of silicon carbide lie in its growth and fabrication maturity relative to diamond \cite{Liu2015, Bracher2015}.  There also exist several commercially viable polytypes which host dozens of candidate defects \cite{Christle2015}.  Among the most attractive defects in SiC are the silicon vacancies $V_\text{Si}$, the divacancies $V_\text{C}V_\text{Si}$, and NV centers \cite{Hain2014, Widmann2014, Carter2015, VonBardeleben2015, Fuchs2015, Soykal2016, Seo2016, Christle2017, Roland2018}.  There are two silicon vacancies in 4H-SiC, V1, which has a zero-phonon line (ZPL) at 862 nm, and V2, which has a ZPL at 916 nm \cite{Sorman2000, Wagner2000, Janzen2009}.  We focus on the negatively charged V2, which has room temperature spin coherence (spin echo $T_2$ of $\sim$100 $\mu$s) \cite{Simin2017, Widmann2014, Carter2015}, like NV centers in diamond, but differentiates itself by having a spin 3/2 ground state, instead of spin 1 \cite{Widmann2014, Carter2015, Aharonovich2016, Soykal2017}.

In this paper, we measure the optical properties of single V2 defects at low temperature and develop a theoretical fine structure model that reveals its complete optical and spin polarization properties for the first time. We perform high-resolution photoluminescence excitation (PLE) measurements as well as time dependent experiments under resonant optical and radio frequency (RF) excitation. The PLE spectra of single defects show two sharp transitions that correspond to the $m_s=\pm 3/2$ and $m_s=\pm 1/2$ spin states. Time-resolved PL measurements under resonant and non-resonant laser excitation, combined with the theoretical model, provide a quantitative understanding of the intersystem crossing and the spin polarization dynamics. These results provide the foundation for optical measurement and control of the V2 defect.

\section{II. Experimental Methods}

\begin{figure}
	\includegraphics{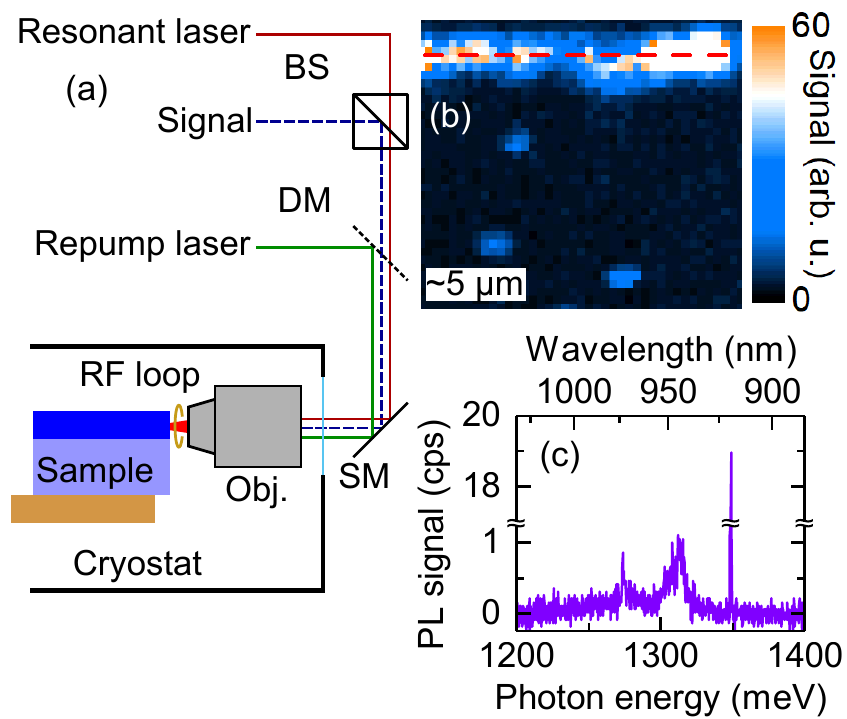}
	\caption{Confocal spectroscopy setup and defect maps. (a) The sample and objective lens (Obj.) are mounted inside the cryostat.  The sample is mounted on edge relative to the optical axis.  The objective is focused inside the epitaxial layer (darker blue).  Two excitation lasers are used.  The resonant laser passes through a 90:10 reflect:transmit non-polarized beam splitter (BS), then is combined with the repump laser by an 850 nm dichroic mirror (DM).  The lasers are then both sent into the objective by a two-axis steering mirror (SM) through a $4f$ lens system (not pictured).  A simple wire loop around the focal spot applies the RF magnetic field. The emitted light travels back through the setup and is reflected off the BS.  (b) Typical raster scan of PL reveals single V2 defects where the PL signal intensity at 916 nm is mapped with a color scale.  The red dashed line denotes the surface of the epitaxial layer. The scale bar is the white rectangle in the lower left, which represents about $\sim$5 $\mu$m. (c) Typical PL spectrum of a V2 defect.}
	\label{experiment_setup}
\end{figure}

To study the ZPL of single V2 defects, we perform high-resolution optical spectroscopy and cryogenic confocal microscopy.  Experimental details are shown in Fig.~\ref{experiment_setup}(a).  The sample is a 18 $\mu$m epitaxial film of 4H-SiC with an n-doping level of $3\times 10^{14}$ cm$^{-3}$, grown by chemical vapor deposition on an n-doped substrate, and then irradiated with $10^{12}$ cm$^{-2}$ 2 MeV electrons \cite{Myers-Ward2014}.  The sample is mounted in a closed-cycle cryostat at $\sim$4.9 K with the c-axis oriented nearly perpendicular \footnote{due to 4$^\circ$ wafer miscut} to the optical axis, such that optical excitation and collection are from the cleaved edge of the sample. This geometry is used because the V2 optical dipole moment is oriented along the c-axis, and we find that the ZPL emission is much stronger from the edge. There are two sources of excitation light: a titanium:sapphire laser at 745 nm, used for photoluminesence (PL) spectra and to reverse resonant bleaching, and an external-cavity diode laser tunable around 916 nm, used for all resonant excitation experiments.  Both lasers are pulsed using acousto-optic modulators.  The two beams are filtered using 850 nm and 925 nm short pass filters, respectively, and are combined using an 850 nm dichroic mirror.  Both beams are focused on the same spot by a 0.75NA objective and are polarized parallel to the c-axis at the sample.  The emitted light, and reflected laser light, is separated from the excitation path using a 90:10 non-polarizing beam splitter and collected by a single-mode fiber that is sent to a detector.

The optical setup is slightly modified depending on the measurement.  For PL measurements, an 850 nm long pass filter is used, and the detector is a silicon CCD spectrometer.  For PLE and other resonant experiments including photobleaching and spin dynamics, a 937 nm long pass filter removes the excitation lasers and either a silicon photon-counting avalanche photodiode or superconducting nanowire detects signal in the phonon sideband (PSB) \footnote{The superconducting nanowire detector increases the signal by a factor of six and the dark counts by a factor of five, approximately, compared to the avalanche photodiode.}.  A radio-frequency (RF) magnetic field for spin-controlled measurements is produced by shorting the inner conductor of a coax cable to the outer conductor, and the loop is stretched over the sample edge so that the RF magnetic field is largely perpendicular to the c-axis.  All experiments are performed at zero applied magnetic field. 

\section{III. Results}

\subsection{A. Single defect emission}
Single defects are located by raster scanning a 745 nm laser and collecting PL to find defects and identify them as V1 or V2 by their ZPL.  The map of one region of the epitaxial layer is shown in Fig.~\ref{experiment_setup}(b), where only emission at the V2 ZPL is plotted. The sample surface is indicated by the red dashed line, which shows bright emission of unknown origin that is not clearly associated with Si vacancies. Below the surface, within the epitaxial layer, there are three bright spots that we attribute to single V2 defects. While this measurement does not definitively indentify single emitters, the density is so low that it is unlikely that two defects are at the same location.  PLE measurements, presented later, indicate that these spots are single defects.  There are about five times as many V1 defects as V2 defects in the same region.  This difference may be due to electron irradiation on the Si-face, consistent with previous studies that the V2 ZPL is significantly dimmer after Si-face irradiation than after C-face irradiation \cite{Sullivan2006}. No defect emission is measured in the substrate, which is likely due to the high n-doping quenching emission.

A single V2 defect has a sharp ZPL and a wide PSB.  A typical PL spectrum for a V2 defect is shown in Fig.~\ref{experiment_setup}(c).  The ZPL at 1352 meV is the strongest, sharpest feature of the PL spectrum, but there are also two broad peaks in the PSB before the CCD response falls off around 1000 nm (1240 meV).  The upper bound on the Debye-Waller factor (DWF), the ratio of the light emitted in the ZPL divided by all of the light emitted by the defect, is about 30\% based on integrating the ZPL and the entire spectrum of Fig.~\ref{experiment_setup}(c). The actual DWF is likely much smaller due to much of the PSB being at longer wavelengths where the Si CCD response is poor.

\begin{figure}
	\includegraphics{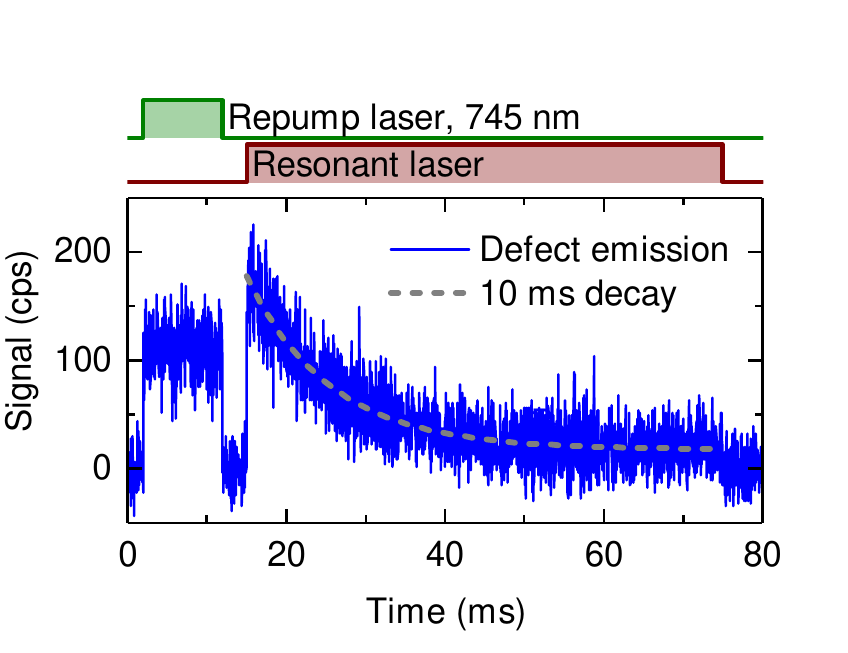}
	\caption{Bleaching measurement.  (top) The pulse sequence for the measurement: 10 ms of 1 mW 745 nm light followed by 60 ms of 10 $\mu$W resonant light.  (bottom) Time trace of the PL from a single defect.  The PL is constant at the millisecond timescale under 745 nm illumination.  During the resonant pulse, the PL from the defect decays away exponentially with a time scale of 10 ms.}
	\label{bleaching}
\end{figure}

Optical excitation resonant with the V2 ZPL induces photobleaching of the defect PL that can be recovered by off-resonant excitation.  The photobleaching signal decay is shown in Fig.~\ref{bleaching} in which a 1 mW, 10 ms, 745 nm laser pulse is followed by a 10 $\mu$W, 60 ms, resonant laser pulse.  Under 10 $\mu$W of resonant illumination, the signal from the defect drops exponentially with a 10 ms time constant.  The signal recovers very quickly under off resonant excitation at 745 nm, faster than the 30 $\mu$s resolution, so it is used as a repump pulse to keep the defect photoactive.  This high repump power, 1 mW, is used to demonstrate the constant PL signal on the same scale as the resonant signal.  In general, much less power and pulse width are needed to repump the defects effectively, and a power of about 100 $\mu$W and a pulse width as small as 200 ns are used for all other resonant experiments.  This photobleaching effect is seen in all V2 defects measured, and the bleaching time constant is generally in the tens of milliseconds.  We suspect that this photobleaching is related to photoionization of the V2$^-$ charge state similar to that in NV$^-$ centers in diamond \cite{Waldherr2011, Beha2012, Siyushev2013}, although there has been no reported observation or identification of the ZPL of the V2$^0$ or V2$^{2-}$ charge states.  The charge state of both the $V_\text{Si}$ and $V_\text{C}V_\text{Si}$ in SiC have previously been shown to be affected by excitation with different wavelengths \cite{Wolfowicz2017, Golter2017}. Another possible mechanism for this photobleaching is optical spin pumping, but we will show that this occurs on much shorter timescales, and the signal can be recovered using RF excitation.

\subsection{B. Photoluminescence Excitation Spectra}

The optical spectrum of the V2 ZPL is measured with high resolution PLE.  The resonant laser is tuned across the ZPL of a defect while the laser frequency is measured by a high-resolution wavemeter.  To prevent bleaching, the pulse sequence consists of an 8 $\mu$s repump pulse followed by a 40 $\mu$s resonant pulse with a 1 $\mu$s delay between the pulses. The emitted light signal is only counted during the resonant pulse. Without the repump pulse, no signal is observed.

\begin{figure}
	\includegraphics{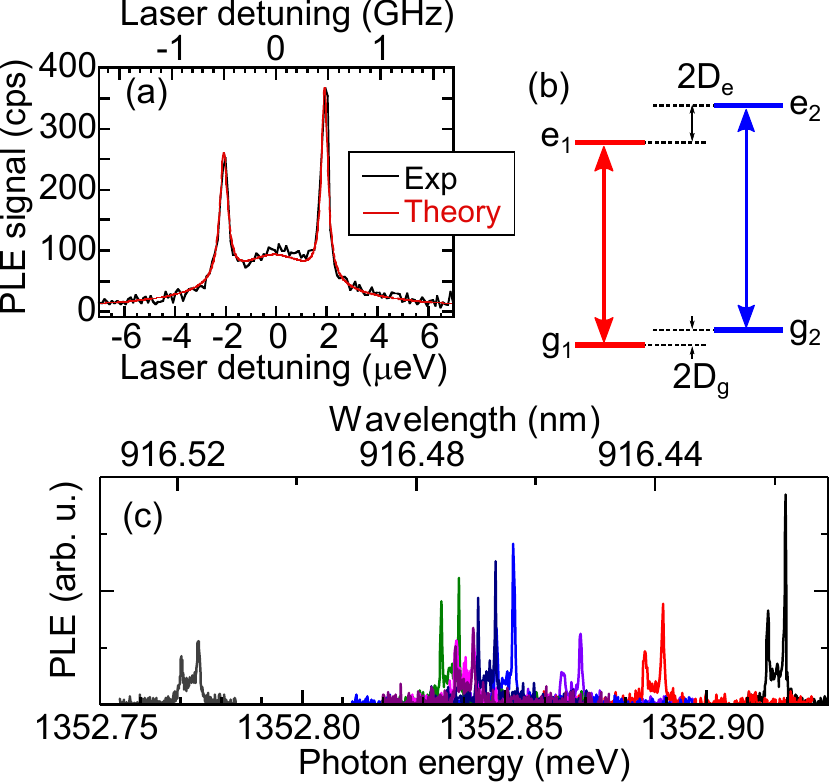}
	\caption{PLE spectra of single defects.  (a) The experimental and theoretical PLE spectrum of a typical single V2 defect showing two sharp peaks and one broad one.  The frequency axis is offset by $1352.8461$ meV and the resonant laser power is 4 $\mu$W.  The two sharp peaks, referred to as ``red'' or ``blue'' transitions, are associated with optical transitions of either the $\pm3/2$ or $\pm1/2$ states.  (b) Simplified energy level diagram showing the red and blue transitions between the ground (g) and excited (e) states. (c) The PLE spectra of nine different defects.  They all have the same overall structure, two sharp peaks and a weak broad peak.}
	\label{simple_spectra}
\end{figure}

The PLE from the V2 defect has two sharp lines consistent with the spin sublevels of the ground (GS) and excited states (ES).  A typical PLE spectrum and energy level schematic is shown in Fig.~\ref{simple_spectra}(a, b).  The two sharp peaks arise from nominally spin-conserving optical transitions between the ground and excited states.  At zero magnetic field, the spin-3/2 ground state is split into two doubly-degenerate sublevels, $m_s{=}\pm 1/2$ and $m_s{=}\pm 3/2$, separated by a zero field splitting (ZFS) of 70 MHz due to spin-spin interactions among the active electrons of the defect \cite{Soykal2016,Carter2015,Mizuochi2002,Mizuochi2003}. Similarly, the spin-3/2 excited state is also split with a ZFS of 1 GHz \cite{Tarasenko2018}. Although the ground and excited state ZFSs have been previously shown to have the same sign \cite{Simin2016}, the overall sign is still debated. We initially classify these sharp peaks as the ``red'' and ``blue'' peaks on the left and right, due to their respective energies, respectively, with corresponding ground ($g_i$) and excited ($e_i$) states. Later, we identify these red and blue peaks to be $m_s = \pm 3/2$ and $\pm 1/2$ spin states, respectively, based on their distinct spin polarization properties. The blue peak is slightly stronger and narrower than the red peak, with full-widths at half-maxima (FWHMs) of 65 MHz and 80 MHz, respectively, which is not much larger than the radiative limit of the linewidth of about 24 MHz (from a fluorescence lifetime of 6.1 ns) \cite{Hain2014}. The splitting between the red and blue transitions is 1.03 GHz, which is consistent with the sum of the ground and excited state ZFSs.

Nine other defects were measured with PLE at a different region of the sample, see Fig.~\ref{simple_spectra}(c).  The blue peak of the defects has an ensemble average of 1352.85 meV and a standard deviation of 40 $\mu$eV (10 GHz), but the splitting between the two peaks has an ensemble average of 1.03 GHz with a standard deviation of 10 MHz, limited by stability of the wavemeter used to measure the excitation frequency. The blue and red peaks have FWHM averages of 170 and 200 MHz, respectively, but with relatively large standard deviations of about 100 MHz.  No spectral diffusion at timescales longer than a measurement (on the order of minutes) was observed. These observations are consistent with low strain and electric field sensitivity of the ZFS \cite{Soykal2017}, and also indicate small ground and excited state permanent dipole moments as expected from symmetrically near-$T_d$ charge distributions \cite{Soykal2016}.

\begin{figure*}
	\includegraphics{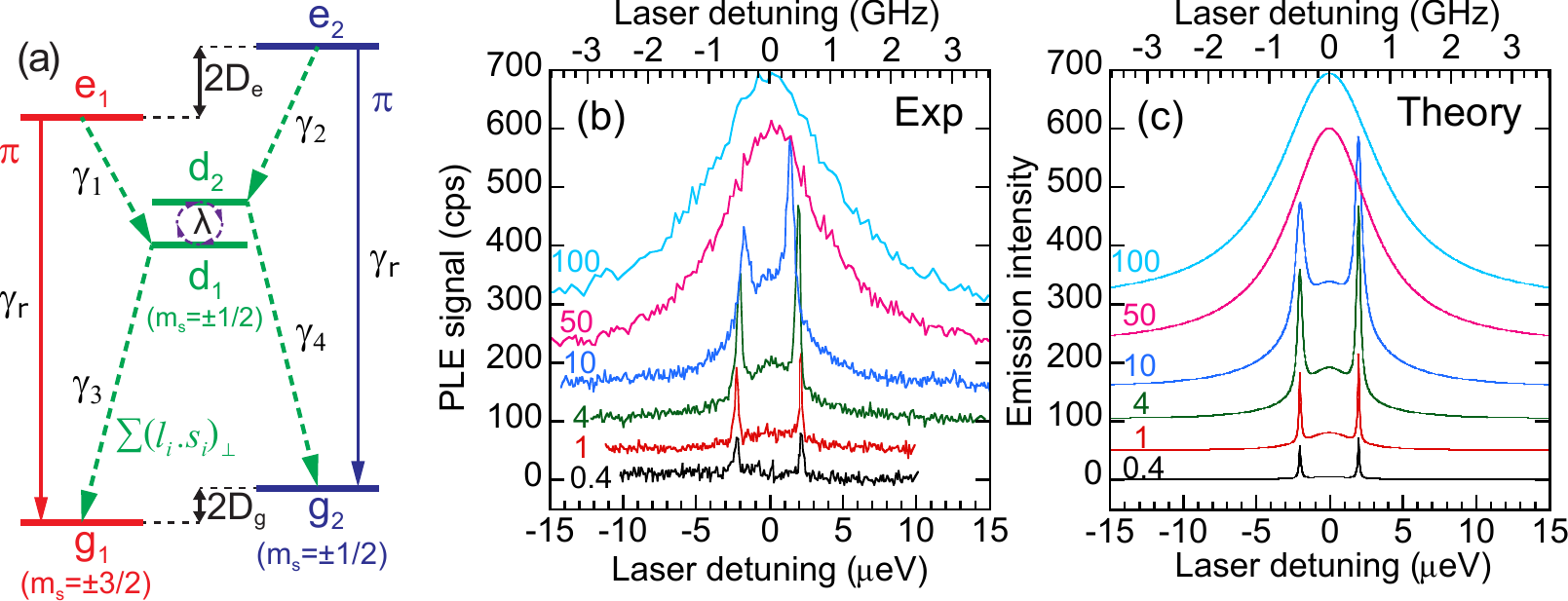}
	\caption{(a) Detailed energy level diagram of V2, showing the red and blue optical transitions and the non-radiative decay channels through the doublets, $d_1$ and $d_2$. (b) Single V2 defect's PLE for a series of resonant laser powers (in $\mu$W). Spectra are offset from each other by 50 cps for clarity, and the laser detuning is relative to 1352.8461 meV. (c) Theoretical PLE spectra for a series of laser powers (in $\mu$W).}
	\label{ple_power_dep}
\end{figure*}

The detailed fine structure of the V2 defect is shown in Fig.~\ref{ple_power_dep}(a) in the absence of any magnetic field. The $m_s{=}\pm 3/2$ and $m_s{=}\pm 1/2$ channels have the same orbital dipole moment and, therefore, have the same radiative decay rates from the ES to the GS. The spin polarization properties and the unique features in the PLE spectra are governed by the phonon-assisted non-radiative inter-system crossing (ISC) processes that are due to direct spin-orbit couplings (DSO) of the $E$ symmetry doublet state (DS) to the $A$ symmetry GS and ES. This leads to four primary ISC rates that need to be determined in order to fully understand the optical properties of this defect. The non-radiative rates that are associated with the decay from the ES to the DS (entry channel) are labeled as $\gamma_1$ and $\gamma_2$ for each spin channel. Similarly, the decay rates from the DS back to the GS (exit channel) are labeled as $\gamma_3$ and $\gamma_4$. We obtain the relationship between different spin channels within each entry and exit channels by evaluating the corresponding DSO coupling matrix elements, e.g. $\gamma_{i}=2\pi|\langle\psi_i^{DS}|\sum_j l_j.s_j|\psi_i^{GS}\rangle|^2/\hbar$ within the symmetry adapted wave-functions basis of each $j$ active electrons \cite{Soykal2016}. In addition, we find that the entry channel ISC rates are ${~{\approx} 2.65}$ times faster than the exit channels taking the relative energy of the DS with respect to GS and ES into account \cite{Soykal2016}. As a result, we obtain the following relationship among all the ISC rates,
\begin{equation}
\gamma_{1}=3\,\gamma_{2}=2.65\, \gamma_{3}=7.95\, \gamma_{4},\label{1}
\end{equation}
and show that $m_s{=}\pm 3/2$ spin states always have 3 times faster non-radiative decay rate than the $m_s{=}\pm 1/2$ states.

The PLE spectra as a function of resonant laser power are shown in Fig.~\ref{ple_power_dep}(b).  At lower resonant laser powers, $P_\text{res}$, we observe two small sharp peaks.  At higher powers there is a broad central peak that dominates the spectrum at the highest powers.  These sharp peaks in PLE spectra can only occur in the presence of spin relaxation that limits the effects of optical pumping. For instance, resonant excitation of the red transition will eventually lead to the defect transitioning through the ISC to $g_2$, a dark state under red excitation. Similarly, resonant optical excitation of the blue transition eventually leads to optical pumping into $g_1$. Therefore, to overcome optical pumping and observe a measurable signal during PLE, the ground state spin must relax reasonably quickly or the laser must be able to excite the opposite transition. Some effective spin relaxation occurs from the periodic repump pulses, but for moderate to high resonant powers, the detuned excitation of the opposite transition plays the biggest role.

The PLE spectra are accurately reproduced in Fig.~\ref{ple_power_dep}(c) by using the theoretical fine structure model of this defect shown in Fig.~\ref{ple_power_dep}(a). In order to obtain the PLE signal theoretically, the steady state excited state populations are calculated from the following Lindblad master equation,
\begin{align}
\frac{d\rho}{dt}=&-\frac{i}{\hbar}[H_0,\rho]+\gamma_r\sum_{i=1}^{2}L(M_r^i)+\sum_{i=1}^{4}\gamma_i L(M_I^i)\nonumber\\
&+\gamma_R\sum_{i=1}^{2}L(M_R)+\gamma_s L(M_s).\label{2}
\end{align}
in which the various decay and decoherence processes of the defect are represented by the Lindblad super-operators, i.e. $L(A)=A\rho A^\dagger-\left\{A^\dagger A,\rho\right\}/2$ for any operator $A$. The fine structure Hamiltonian of the V2 defect (in the absence of any magnetic field) is given by
\begin{align}
H_0=&\left(D_g-D_e+\delta_{L}\right)\left(|g_1\rangle\langle g_1|-|e_1\rangle\langle e_1|\right)/2\nonumber\\
&-\left(D_g-D_e-\delta_{L}\right)\left(|g_2\rangle\langle g_2|-|e_2\rangle\langle e_2|\right)/2\nonumber\\
&+\lambda\left(|d_1\rangle\langle d_2|+|d_2\rangle\langle d_1|\right)\nonumber\\
&+\Omega_L\left(|g_1\rangle\langle e_1|+|g_2\rangle\langle e_2|+c.c.\right),\label{3}
\end{align}
in the rotating frame of the PLE laser with a frequency detuned by $\delta_L$ from the ZPL and a Rabi amplitude of $\Omega_L$. Our model does not assume the sign of the GS and ES ZFS but only that they have the same sign, which has previously been demonstrated \cite{Dyakonov2016}. The ZFS sign and all the relevant decay rates (see Table \ref{table1}) are obtained later by applying this fine structure model to the time resolved spin polarization and readout measurements, which will be explained later. The ZFS of the GS and ES are $2D_g$ and $2D_e$, respectively. We note that the $E$ doublet with doubly degenerate $d_1$ and $d_2$ spin states is the only doublet that is allowed to couple to both GS and ES via ISC. Any additional spin-mixing effects stemming from the DSO couplings of this doublet with the remaining four energetically close doublets ($A_1, A_2, 2E$) \cite{Soykal2016} are included in the parameter $\lambda$. The radiative decays are given by $M_r^i=|g_i\rangle\langle e_i|$ in Eq.~\ref{2} and governed by the same radiative decay rate $\gamma_r$. On the other hand, the 4 non-radiative ISC decay processes involving the doublet state are given by $M_I^1=|d_1\rangle\langle e_1|$, $M_I^2=|d_2\rangle\langle e_2|$, $M_I^3=|g_1\rangle\langle d_1|$, and $M_I^4=|g_2\rangle\langle d_2|$ with rates $\gamma_{i}$. We also include the spin relaxation rate $\gamma_R$ of the ground state defined in $M_R=|g_1\rangle\langle g_2|+c.c.$ due to the repump laser and the intrinsic spin dephasing rate $\gamma_s$ of the doublet states, i.e. $M_s=|d_1\rangle\langle d_1|-|d_2\rangle\langle d_2|$. $\gamma_s$ and $\lambda$ are found to be large enough that the doublet states are completely mixed. Dephasing of the GS spin is not included as it is typically on a longer time scale than the dynamics considered here, and it should not affect the population dynamics.

The amplitude difference in the sharp peaks can be understood from the different ISC rates of the $m_s{=}\pm 3/2$ and $m_s{=}\pm 1/2$ spin channels (see Eq.~\ref{1}). The faster ISC rate of the $m_s{=}\pm 3/2$ excited states causes more population to be removed non-radiatively from these states per optical cycle compared to the $m_s{=}\pm 1/2$ states. Therefore, the $m_s{=}\pm 3/2$ excited states have smaller steady state populations than the $m_s{=}\pm 1/2$ excited states, when the laser is on or near resonant with their corresponding transitions, leading to a reduced PLE signal from that peak. As a result, we identify the smaller sharp peak as the $m_s{=}\pm 3/2$ states leading to a negative ZFS ($D<0$) for the ES and the GS. This identification will be corroborated by our time-dependent spin polarization and readout measurements. As the $P_\text{res}$ is increased, as expected by the model, the sharp peaks get stronger and broader while the amplitude difference between them increases before finally saturating. In addition, a broad peak, that increases in amplitude and linewidth with increasing power, appears. This broad peak, which is prominent only at higher $P_\text{res}$, is due to off-resonant excitation of both the red and blue transitions, which is strongest in between the two transitions.  Symmetric excitation of both transitions prevents optical pumping.  This leads to a stationary center frequency that remains halfway between the sharp peaks and a peak area that increases linearly before beginning to saturate at about 50 $\mu$W. The sharp peaks also appear to move toward the center at higher powers for reasons that are not yet clear. One possibility is that strong resonant driving modifies the charge environment surrounding the defect, causing electric field induced changes to the ZFS.

\subsection{C. Time-resolved spin dynamics}

\begin{figure}
	\includegraphics{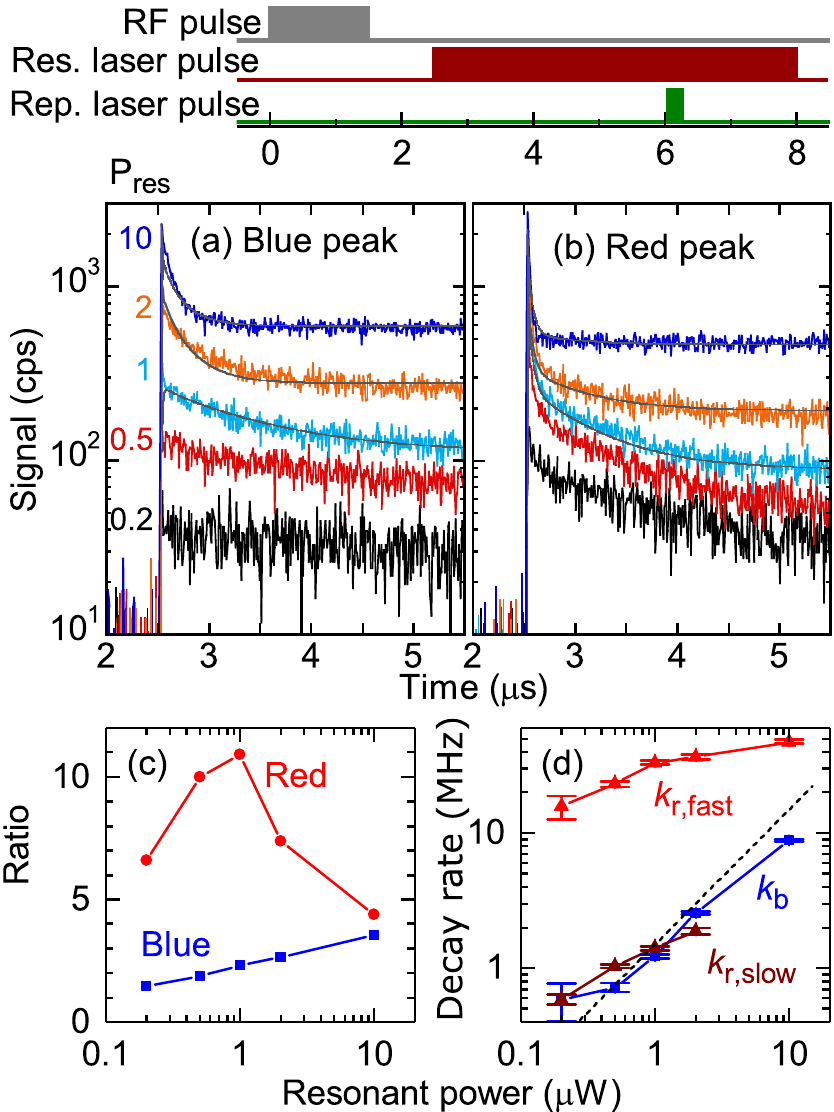}
	\caption{PL decay due to optical spin pumping and ISC. Above: Pulse sequence for time-resolved PL measurements. The RF pulse equalizes spin populations. (a,b) Time resolved PL from the PSB while driving the (a) blue and (b) red transitions for a series of resonant laser powers (in $\mu$W). Grey curves superimposed on the experimental data are from the theoretical model. (c) The ratio of the initial peak of the signal to the steady-state signal as a function of $P_\text{res}$.  (d) Decay rates from fits to the measurements above.  The error bars are approximated using the diagonal elements of the covariance matrix.  The dashed line is a linear relationship as a guide to the eye.  The decay under blue excitation is described well by a single exponential decay rate $k_\text{b}$, while the decay under red excitation requires two rates, $k_\text{r,fast}$, $k_\text{r,slow}$.}
	\label{spin_polarization_rates}
\end{figure}

The dynamics of this system under laser excitation are measured using RF pulses and time-resolved PL.  The integrated PSB emission is measured as a function of time and for a series of resonant laser powers, shown in Fig.~\ref{spin_polarization_rates}(a,b).  A 1.5 $\mu$s, 72.5 MHz RF pulse, resonant with the ground state ZFS \footnote{A small shift from the 70 MHz ZFS resonance was measured for reasons that are unclear.} is applied before the resonant laser pulse to equalize the spin polarization left over from the previous resonant laser pulse.  The RF pulse length and power are adjusted with the intent to equalize the spin populations when Rabi oscillations are damped out, but this may not be entirely the case. Regardless, the RF pulse at least partially equalizes spin states and allows observation of the PL decay.  The resonant laser pulse is locked by the wavemeter to either the red or blue transition at every given $P_\text{res}$.  A short repump pulse long after the dynamics have reached their steady-states prevents the defect from bleaching.  Both excitation laser frequencies initially show strong PSB emission that decays very rapidly due to the ISC and optical spin polarization. The PL does not decay to zero due to off-resonant excitation of the other ground state, giving rise to steady state emission. Spin relaxation should also contribute to the steady state emission at low powers.

Driving the two transitions resonantly leads to very different time dependences. Decay rate results are shown in Fig.~\ref{spin_polarization_rates}(d).  Under blue excitation after the RF pulse, the signal can be fit very well with a single exponential, giving a single polarization rate $k_\text{b}$.  Under red excitation after the RF pulse, however, the signal requires a biexponential fit, which yields to two polarization rates, $k_\text{r,fast}$ and $k_\text{r,slow}$.  As expected, the spin polarization rates increase with increasing power.  At 10 $\mu$W red excitation, however, only the fast rate is observed.

The decay rates as a function of optical power help give an intuitive picture for the spin-dependent ISC coupling constants and further support the theoretically obtained relationship (Eq.~\ref{1}) among these rates. These coupling constants determine optically-pumped ground state spin polarization. As shown in Fig.~\ref{ple_power_dep}(a), for a given resonant excitation, blue or red, there are two primary parameters, the excited state-to-intermediate state decay rate ($\gamma_2$ or $\gamma_1$), and the intermediate state-to-opposite ground state decay rate ($\gamma_3$ or $\gamma_4$).  The former determines how fast the defect can leave the resonant radiative transition cycle (related to how well optical spin readout works), and latter determines how likely the defect is to relax to the other dark ground state, leaving the radiative transition cycle (related to the rate of optical pumping).
The different time dependences of red and blue excitations can be explained very well using the model in Fig.~\ref{ple_power_dep}(a) and the relationships in Eq.~\ref{1}. PL decay curves obtained from this model are plotted in Fig.~\ref{spin_polarization_rates}(a,b), showing very good agreement with experiment. For the red excitation, as the $\gamma_1$ decay rate is faster than both the $\gamma_3$ and $\gamma_4$, the defect spends a significant fraction of a cycle in the DS, which leads to a quick decrease in luminescence (due to population moving into the DS) that occurs ``in series'' with the optical pumping. Therefore, the decay of red excitation exhibits a biexponential time dependence. On the other hand, for the blue excitation, since the $\gamma_2$ decay rate is comparable to the $\gamma_3$ decay rate, the spin polarizes to the other dark ground state nearly as quickly as it decays into the DS, and so only one rate is observed. These results significantly strengthen the assignment of the red and blue transitions to the $m_s{=\pm}3/2$ and $m_s{=\pm}1/2$ states, respectively.

\begin{figure*}
	\includegraphics[width=17 cm]{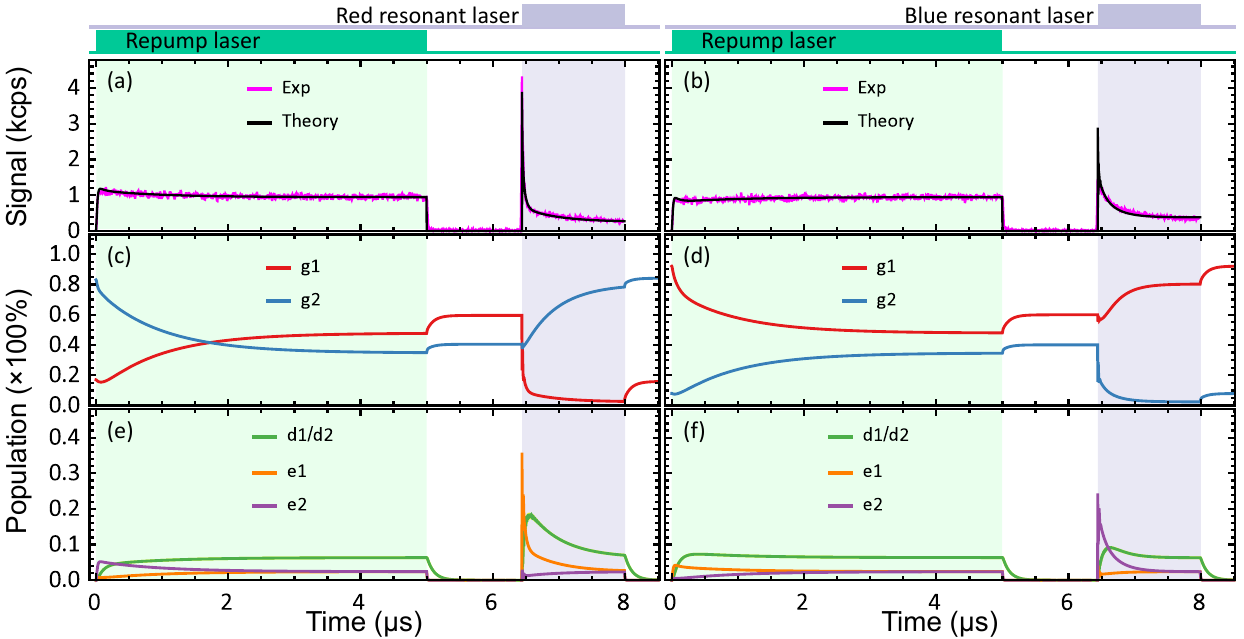}
	\caption{Spin dynamics for pulse sequences of a non-resonant repump pulse ($1$mW) followed by a red (left column) or blue (right column) resonant pulse ($5\mu$W). (a, b) Measured and modeled PL signal from a defect as a function of time. (c, d) Calculated ground state populations during the sequence extracted from the model fit to the experimental signal above. (e, f) Calculated excited state and doublet state populations during the sequence.}	\label{off_res_polarization}
\end{figure*}

Understanding spin dynamics under non-resonant laser excitation is also valuable, due to its common use in polarizing and reading out V2 and its use in preventing bleaching of emission. As shown in Fig.~\ref{off_res_polarization}, a combination of a non-resonant/repump (745 nm) pulse and a resonant pulse on the red or blue transition are used to study these dynamics.

Both pulses of the sequence serve to readout and polarize the ground state spin, but the resonant laser acts more selectively. In the red pulse sequence, just before the repump pulse turns on at $t = 0$, the defect has primarily been optically pumped into $g_2$ by the red-tuned laser. Just after the repump pulse begins which excites both transitions, the PL signal in Fig.~\ref{off_res_polarization}(a) starts at a slightly higher value than the steady state signal at $t=5$ $\mu$s. For the blue pulse sequence, the defect has primarily been pumped into $g_1$, and the PL signal in Fig.~\ref{off_res_polarization}(b) starts at a slightly lower value than the steady state. This behavior is consistent with the results from Fig.~\ref{spin_polarization_rates}, which show that the non-radiative ISC processes are faster for $e_1$ than $e_2$, so more photons are collected when starting in $g_2$ compared to starting in $g_1$. The difference in PL signal, integrating over the $1/e$ decay time (1.2 $\mu$s), is 18\%. This contrast is much larger than the 2\% seen in room temperature single defect ODMR \cite{Widmann2014}, but is still relatively small compared to resonant readout because contrast is only due to the differences in the non-radiative ISC processes. It is also possible that there may be other luminescent defects within the laser spot that contribute to the signal, artificially lowering the contrast.

We also apply the theoretical model of Eq.~\ref{2} and Eq.~\ref{3} to these sequences of resonant and non-resonant pulses. The comparison to experiment for these sequences is very valuable for determining the system parameters and for visualizing the populations that cannot be separately measured experimentally. The theoretical PL signal in Fig.~\ref{off_res_polarization}(a,b) is calculated by evaluating the time dependent excited state populations of $e_1$ and $e_2$ as a function of two decay rates, $\gamma_r$ and $\gamma_1$, where the ratio between all the ISC rates is kept fixed according to Eq.~\ref{1}. Fitting these solutions to the experimental data reveals all the previously unknown radiative and non-radiative decay rates of the V2 defect that are shown in Table \ref{table1}. In addition, these calculated rates result in an overall fluorescence lifetime of $5.82$ ns (i.e. $1/\gamma_F=1/\gamma_r+4/\gamma_1$) in good agreement with previous measurements \cite{Hain2014}.

We also conclude that both the GS and the ES ZFS are negative ($D_g<0$, $D_e<0$) as the red excitation with higher ISC rate corresponds to the $m_s=\pm 3/2$ spin states (see Eq.~\ref{1}). The population of each state is plotted in Fig.~\ref{off_res_polarization}(c, e) and Fig.~\ref{off_res_polarization}(d, f) for the red and blue resonant laser, respectively. During the repump laser, the initial $g_1$ or $g_2$ population is pumped toward a modest $g_1$ polarization, which reaches a population of 0.59 after the laser is off and all population is returned to the GS. This polarization occurs because of faster decay from the doublets to $g_1$ than $g_2$.  During the resonant pulse, polarization occurs very rapidly toward $g_1$ or $g_2$, with the $g_1$ ($g_2$) population reaching 0.92 (0.84) for blue (red) excitation after the pulse is turned off. These values are limited by off-resonant excitation of the opposite transition. Based on this model, using lower resonant laser powers of ${\approx}15$ nW and pulse lengths of $4$ $\mu$s, the initialization fidelity into $g_1$ ($g_2$) can be $0.9996$ ($0.9993$) for blue (red) excitation.
\setlength{\tabcolsep}{15pt}
\renewcommand{\arraystretch}{1.}
\begin{table}[!htp]
\centering
  \begin{tabular}{cc}
	 Transition & Lifetime (ns)\\
	\hline\hline
		$1/\gamma_r$ & $8.2$ \\
		$1/\gamma_1$ & $26.7$\\
		$1/\gamma_2$ & $80$\\
		$1/\gamma_3$ & $70.6$\\
		$1/\gamma_4$ & $211.9$
  \end{tabular}\caption{Radiative $\gamma_r$, and non-radiative decay rates $\gamma_i$ of the V2 defect at 5 $\mu$W resonant laser power.}\label{table1}
\end{table}

\section{IV. Discussion and Conclusion}

The results of this article are relevant to applications in quantum technology, particularly for quantum networks \cite{Kimble2008, Awschalom2018}, which require fast, high fidelity initialization, efficient spin readout, stable, homogeneous emission into the ZPL, and the proper energy level structure. As just stated, the initialization fidelity for V2 looks very promising under the right excitation conditions. Fast initialization requires a rapid change in the spin state under optical excitation, which is true for both transitions.

Aside from efficient photon collection, the key to spin readout is to preserve the spin state under optical excitation, opposite the requirement for initialization. Readout of the spin state is better for the blue transition since the ISC rate to the doublet is 3 times slower than for the red transition, meaning more photons can be emitted before the defect changes spin state. The blue transition, however, does not operate very well as a cycling transition since the ISC is still relatively strong. When starting in the $g_2$ ($m_s = \pm1/2$) state, we estimate that about 8 photons should be emitted when driving the blue transition before the emission decays to $1/e$ of its initial value. This behavior is in contrast to the $S=1$ divacancy in SiC and NV$^-$ in diamond, in which an ES for the $m_s=0$ GS has a weak ISC for low strain-induced mixing of excited states \cite{Robledo2011, Christle2017}, giving long PL decay times for efficient spin readout. Furthermore, for the V2 $\rm{V_{Si}}$ our model shows that the long lifetime of the DS ($\gamma_3 < \gamma_r,$ and $\gamma_4 < \gamma_r$) limits the overall brightness by causing some population to become trapped in the DS (see Fig. 6(e,f)). Integration of these defects into cavities should improve readout by increasing the collection efficiency and by Purcell enhancement of the radiative transitions relative to the ISC. Cavity enhancement of different optical transitions in semiconductor QDs has been successful in improving the fidelity of both readout and initialization \cite{Carter2013, Sun2016, Sun2018}.

Obtaining homogeneous emission into the ZPL depends on having little emission into the PSB, little variation in emission energy between defects, and little spectral diffusion of individual defects. More than 70\% of the emission from V2 is in the PSB, but this is still likely better than 97\% for the NV center \cite{Alkauskas2014}. The inhomogeneous variation in emission energy, which we attribute to local variations in strain and electric field, and the measured linewidths of single defects, broadened by fluctuations in the environment, are roughly comparable to that of high quality diamond NV center samples \cite{Santori2006, Togan2010}, but we expect these may be improved using different irradiation conditions and annealing.

The energy level structure required depends on the application. For spin-photon entanglement, $\Lambda$ energy level systems are often used for polarization \cite{Togan2010} or frequency \cite{Gao2012} encoding of the photon, but entanglement can also be achieved without a $\Lambda$ system using time-bin encoding of photons, as has been demonstrated with NV centers in diamond \cite{Bernien2013, Hensen2015}.  This time bin approach appears feasible with the current V2 energy level structure. The generation of entangled photon states (e.g. cluster states or GHZ states) has been studied for both $\Lambda$ systems \cite{Rao2015} and double two level systems with orthogonal polarizations \cite{Lindner2009, Schwartz2016, Soykal2016}, both of which would require additional additional work to achieve with V2.

The resonant optical experiments presented here provide the essential foundation for an interface between the coherent GS spin system of the V2 $\rm{V_{Si}}$ and photons. The spin system has already been shown to have long coherence times, up to 10s of milliseconds at low temperatures with dynamical decoupling \cite{Simin2017}. The combination of this spin memory with sharp, stable optical transitions, fast initialization, and potentially strong emission into the ZPL make it quite promising as an optically active spin qubit. The ability to integrate this defect into the technologically mature SiC material system is attractive as well.

After submission of this article, a related paper appeared studying resonant optical excitation of the V1 $\rm{V_{Si}}$ \cite{Nagy2018}, with complementary results.

\begin{acknowledgments}
This work was supported by the U.S. Office of Naval Research and the OSD Quantum Sciences and Engineering Program. H.B. Banks acknowledges the National Research Council Research Associate Program.  We also acknowledge Evan Glaser, Joel Grim, Allan Bracker and Dan Gammon for helpful discussions.

H. B. Banks and \"{O}. O. Soykal contributed equally to this manuscript.
\end{acknowledgments}


%

\end{document}